\documentclass[12pt]{article}
\usepackage[utf8]{inputenc}
\usepackage{graphicx}
\usepackage{subcaption}
\usepackage{topcapt}
\usepackage{booktabs}
\usepackage{cite}
\usepackage{url}
\usepackage{hyperref}
\usepackage[margin=1.0in]{geometry}

\title{Origin of Life in a Digital Microcosm}
\author{Nitash C G$^{1,2}$, Thomas LaBar$^{2,3,4}$, Arend Hintze$^{1,2,4,5}$, Christoph Adami$^{2,3,4,6}$}
\date{}

\begin{document}

\maketitle
\begin{center}
$^{1}$ Department of Computer Science \& Engineering\\
$^{2}$ BEACON Center for the Study of Evolution in Action\\
$^{3}$ Department of Microbiology \& Molecular Genetics\\
$^{4}$ Program in Ecology, Evolutionary Biology, and Behavior\\
$^{5}$ Department of Integrative Biology\\
$^{6}$ Department of Physics and Astronomy\\
Michigan State University, East Lansing, MI 48824\\
\end{center}

\begin{abstract}
While all organisms on Earth descend from a common ancestor, there is no consensus on whether the origin of this ancestral self-replicator was a one-off event or whether it was only the final survivor of multiple origins. 
Here we use the digital evolution system Avida to study the origin of self-replicating computer programs. By using a computational system, we avoid many of the uncertainties inherent in any biochemical system of self-replicators (while running the risk of ignoring a fundamental aspect of biochemistry). We generated the exhaustive set of minimal-genome self-replicators and analyzed the network structure of this fitness landscape. We further examined the evolvability of these self-replicators and found that the evolvability of a self-replicator is dependent on its genomic architecture. We studied the differential ability of replicators to take over the population when competed against each other (akin to a primordial-soup model of biogenesis) and found that the probability of a self-replicator out-competing the others is not uniform. Instead, progenitor (most-recent common ancestor) genotypes  are clustered in a small region of the replicator space. Our results demonstrate how computational systems can be used as test systems for hypotheses concerning the origin of life.
\end{abstract}
\section*{Introduction}

There is perhaps no topic in biology more fascinating--and yet more mysterious--than the origin of life. With only one example of organic life to date, we have no way of knowing whether the appearance of life on Earth was an extraordinarily rare event, or it if was a commonplace occurrence that was unavoidable given Earth's chemistry. Were we to replay Earth's history one thousand times~\cite{gould1990wonderful}, how often would it result in a biosphere? And among the cases where life emerged, how different or how similar would the emergent biochemistries be? 

The role of historical contingency has been studied extensively in the evolution of life (see, e.g., \cite{Blountetal2008} and references therein). Here we endeavour to ask an even more fundamental question: What is the role of historical contingency in the origin of life?  The best evidence suggests that the first self-replicators were RNA-based~\cite{gilbert1986origin,robertson2012origins}, although other first self-replicators have been proposed~\cite{leslie2004prebiotic}. Given the large number of uncertainties concerning the possible biochemistry that would lead to the origin of self-replication and life, either on Earth or other planets, researchers have begun to study the process of emergence in an abstract manner. Tools from computer science~\cite{pargellis1996spontaneous,Hutton2002,pargellis2003self,dorn2011monomer,walker2013algorithmic,GreenbaumPargellis2017}, information theory~\cite{walker2014top,adami2015information,adami2015entropy,davies2016hidden}, and statistical physics~\cite{england2013statistical,mathis2015emergence} have been used in an attempt to understand life and its origins at a fundamental level, removed from the peculiarities of any particular chemistry. Investigations along those lines may reveal to us general laws governing the emergence of life that are obscured by the $n=1$ nature of our current evidence, point us to experiments that probe such putative laws, and get us closer to understand the inevitability--or perhaps the elusiveness--of life itself~\cite{cronin2016beyond}.

At the heart of understanding the interplay between historical contingency and the origin of life lies the structure of the fitness landscapes of these first replicators, and how that landscape shapes the biomolecules' subsequent evolution. While the fitness landscapes of some RNA-based genotypes have been mapped~\cite{jimenez2013comprehensive,petrie2014limits} (and other RNA replicators have been evolved experimentally~\cite{mills1967extracellular}), in all such cases evolution already had the chance to shape the landscape for these organisms and ``dictate", as it were, the sequences  most conducive for evolution.

The structure of primordial fitness landscapes, in comparison, is entirely unknown. While we know, for example, that in realistic landscapes highly fit sequences are genetically close to other highly fit sequences (this is the essence of Kauffman's ``Central Massif" hypothesis~\cite{Kauffman1993}, see also~\cite{Ostmanetal2010}), we suspect that this convenient property--which makes fitness landscapes ``traversable"~\cite{Ostmanetal2010}--is an outcome of evolution, in particular the evolution of evolvability. What about primordial landscapes not shaped by evolution? How often are self-replicators in the neighborhood of other self-replicators? Are self-replicators evenly distributed among sequences, or are there (as in the landscapes of evolved sequences) vast areas devoid of self-replicators and rare (genetic) areas that teem with life? Can evolution easily take hold on such primordial landscapes? 

These are fundamental questions, and they are central to our quest to understand life's origins. 
If the fitness landscape consist of isolated fitness networks, as found in some modern RNA fitness landscapes~\cite{jimenez2013comprehensive,petrie2014limits}, then one may expect the effects of historical contingency to be strong, and the future evolution of life to depend on the characteristics of the first replicator. However, if there exist ``neutral networks'' that connect genotypes across the fitness landscape (as found in computational RNA landscapes~\cite{huynen1996smoothness}) then the effect of history may be diminished. Can we learn more about these options?

Recently, we have used the digital evolution platform Avida as a model system to study questions concerning the origin of life~\cite{ofria2009avida}. In Avida, a population of self-replicating computer programs undergo mutation and selection, and are thus undergoing Darwinian evolution explicitly~\cite{pennock2007models}. Because the genomic content required for self-replication is non-trivial, most Avidian genomes are non-viable, in the sense that they cannot form ``colonies" and thus propagate information in time. Thus, viable self-replicators are rare in Avida, with their exact abundance dependent on their information content~\cite{adami2015information,adami2015entropy}. Further work on these rare self-replicators showed that while most of them were evolvable to some degree, their ability to improve in replication speed or evolve complex traits greatly varied~\cite{labar2015does}. Furthermore, the capability of avidian self-replicators to evolve greater complexity was determined by the {\em algorithm} they used for replication, suggesting that the future evolution of life in this digital world would be highly contingent on the original self-replicator~\cite{labar2015evolvability}. However, all of this research was performed without a complete knowledge of the underlying fitness landscape, by sampling billions of sequences of a specific genome-size class, and testing their capacity to self-replicate. 

Sequences used to seed evolution experiments in Avida are usually hand-written~\cite{Adami1998,Adami2006}, for the simple reason that it was assumed that they would be impossible to find by chance. Indeed, a typical hand-written ancestral replicator of length 15 instructions is so rare--were it the only replicator among sequences of that length--that it would take a thousand processors, executing a million sequences per second each in parallel, about 50,000 years of search to find it~\cite{adami2015entropy}. However, it turns out that shorter self-replicators exist in Avida. An exhaustive search of all 11,881,376 sequences of length $L=5$, as well as all 308,915,776 sequences of length $L=6$ previously revealed no self-replicators~\cite{adami2015entropy}. However, in that investigation six replicators of length $L=8$ turned up in a random search of a billion sequences of that length, suggesting that perhaps there are replicators among the 8 billion or so sequences of length $L=7$. 

Here, we confirm that the smallest replicator in Avida must have 8 instructions by testing all $L=7$ sequences, but also report mapping the entirety of the $L=8$ landscape ($26^8\approx 209\times 10^9$ sequences) to investigate the fitness landscape of primordial self-replicators of that length. Mapping all sequences in this space allows us to determine the relatedness of self-replicators and study whether they occur in clusters or evenly in sequence space, all without the usual bias of studying only sequences that are among the ``chosen" already. Of the almost 209 billion possible genomes, we found that precisely 914\footnote{The sequences of all replicators can be downloaded from \url{10.6084/m9.figshare.4551559}.} could undergo self-replication and reproduction, and thus propagate their information forward in time in a noisy environment.

We found that these 914 primordial replicators are not uniformly distributed across genetic space, but instead cluster into two broad groups (discovered earlier in larger self-replicators~\cite{labar2015evolvability}) that form 13 main clusters. By analyzing how these groups (and clusters) evolve, we are able to study how the primordial landscape shapes the evolutionary landscape, and how chance events early in evolutionary history can shape future evolution.

\section*{Methods}

\subsection*{Avida}

We used Avida (version 2.14) as our computational system to study the origin of self-replication. Avida is a digital evolution system in which a population of computer programs compete for the system resources needed to reproduce (see~\cite{ofria2009avida} for a full description of Avida). Each of these programs is self-replicating and consists of a genome of computer instructions that encode for replication. During this asexual reproduction process, mutations can occur, altering the speed at which these programs reproduce. As faster replicators will out-reproduce slower replicators, selection then leads to the spread of faster replicators. Because avidian populations undergo Darwinian evolution, Avida has been used to explore many complex evolutionary processes~\cite{lenski1999genome,adami2000evolution,wilke2001evolution,chow2004adaptive,covert2013experiments,goldsby2014evolutionary,zaman2014coevolution}.

The individual computer programs in Avida are referred to as avidians. They consist of a genome of computer instructions and different containers to store numbers. Each genome has a defined start point and instructions are sequentially executed throughout the avidian's lifetime. Some of these instructions allow the avidian to start the replication process, copy their genome into a new daughter avidian, and divide into two avidians (see~\cite{labar2015evolvability} for the full Avida instruction set). During this replication process, mutations can occur, causing the daughter avidian's genome to differ from its parent. These mutations can have two broad phenotypic outcomes. First, mutations can alter the number of instruction executions required for replication; these mutations can increase or decrease replication speed and thus fitness. Second, the fixation of multiple mutations can lead to the evolution of complex traits in Avida. These traits are the ability to input binary numbers from the Avida environment, perform Boolean calculations on these numbers, and then output the result of those calculations. In the experiments described here, avidians could evolve any of the nine one- and two-input logic functions (Not, Nand, OrNot, And, Or, AndNot, Nor, Xor, and Equals). This is usually referred to as the ``logic-9" environment~\cite{lenski2003evolutionary}.

The ability to perform the above Boolean logic calculations (possess any of these nine traits), increases its bearer's replication speed by increasing the number of genome instructions the bearer can execute per unit of time. The more instructions an avidian can execute during a unit of time, the fewer units of time that are required for self-replication. These units of time are referred to as updates (they are different from generations). During each update, the entire population will execute $30N$ instructions, where $N$ is the current population size. The ability to execute one instruction is called a ``Single Instruction Processing" unit, or SIP. If the population is monoclonal, each avidian will receive, on average, 30 SIPs. However, every avidian also has a {\em merit} which determines how many SIPs they receive per update. The greater the merit, the more SIPs that individual receives. The ability to perform the nine calculations multiply an individual's merit by the following values: Not and Nand: 2, OrNot and And: 4, AndNot and OR: 8, Nor and Xor: 16, and Equals: 32. 

The Avida world consists of a fixed-size toroidal grid of cells. The total number of cells sets the maximum population size. Each cell can be occupied by at most one avidian. After successful reproduction, a new avidian is placed into one of the world's cells. In a well-mixed population, any cell in the population may be chosen. In a population with spatial structure, the new avidian is placed into one of the nine cells neighboring the parent avidian (including the cell occupied by the parent). If there are empty cells available, the new avidian occupies one of these cells. If all possible cells are occupied, a cell is chosen at random, its occupant removed from the population, and the new avidian then occupies this cell. This random removal implements a form of genetic drift in Avida. For the experiments performed here, the population structure was spatial.

\subsection*{Experimental Design}

In order to map the entire Avida fitness landscape, we constructed all $26^{8}\approx 2.09\times10^{11}$ genomes and analyzed whether they could self-replicate. This operation was performed by running these genomes through Avida's {\em Analyze Mode} (described in the Data Analysis section) and checking whether these genomes gave their bearer non-zero fitness, and whether they were {\em viable}. Next, we described the fitness landscape by looking for the presence of genotype clusters among the discovered self-replicators. We constructed a network of the fitness landscape where each genotype is a node and the length between two nodes is the square of the Hamming distance between the genotypes. We also examined the frequency of single instruction motifs (monomers), as well as double instruction motifs (dimers). 

To test the evolvability of the 914 self-replicators, we evolved 10 monoclonal populations of each replicator with 3,600 individuals for $2\times10^{4}$ updates in the logic-9 environment (see above). Point mutations occurred at a rate of $7.5\times 10^{-3}$ mutations per copied instruction, while single-instruction insertion and deletion mutations both occurred at a rate of $5\times 10^{-2}$ mutations per division. At the end of each population's evolution, we analyzed the most abundant genotype from each population.  
  
In order to test the role of historical contingency when the appearance of self-replicators was frequent, we ran experiments where we evolved all 914 self-replicators in the same population (a ``primordial soup" of replicators). In each population, we placed 10 individuals of each self-replicator. The ancestral population then had 9140 individuals and could expand to $10^4$ individuals at maximum capacity. These populations evolved for $5\times 10^{4}$ updates in the logic-9 environment. Mutation rates were the same as in the previous evolvability experiments. This experiment was performed in 200 replicates. To identify the ancestral genotype that outcompeted all of the other genotypes, we isolated the most abundant genotype at the end of the experiment and traced its evolutionary history back to its original ancestor.

\subsection*{Data analysis}

Statistics on different avidians were calculated using Avida's {\em Analyze Mode}. In Analyze Mode, a single genotype is examined in isolation as it executes the instructions in its genome, runs through its life-cycle, and possibly creates an offspring. This confers on experimenters the ability to calculate the fitness for an individual avidian (number per offspring generated per unit time) and examine other characteristics, such as whether it can reproduce perfectly (all offspring are genetically identical to each other and the mother genome) or which traits this avidian possesses. Analyze Mode was also used to calculate quantities such as genome size. Avida's analyze mode code is available along with the entire Avida software at https://github.com/devosoft/avida. 

Across-population means and standard errors were calculated using the NumPy~\cite{van2011numpy} Python software package. The clusters of replicators were rendered using Neato, which is an undirected graph embedder that creates a layout similar to that of Multi-Dimensional Scaling~\cite{Gansner00anopen}. Figures were plotted using the Matplotlib Python package~\cite{hunter2007matplotlib}. 

\section*{Results}

\subsection*{Structure of the Fitness Landscape}

Of the $26^8$ (approximately 209 billion) genomes with 8 instructions, we found 914 that could self-replicate. We also searched for self-replicators with seven-instruction genomes but found none, establishing that $L=8$ is the minimal self-replicator length in Avida. By discovering all self-replicators in this fitness landscape, we can now calculate the precise information content required for self-replication in Avida, using previously-established methods~\cite{adami2015information}, as $-\log_{26}(\frac{914}{26^8}) \approx 5.9$ mers (a ``mer" is a unit of entropy or information, normalized by the number of states that each instruction can take on, see~\cite{Adami2004}). Our previous estimate~\cite{adami2015entropy} of the information content of length-8 replicators, based on finding 8 replicators among a billion random samples, was $5.81\pm0.13$ mers. 

\begin{figure}
    \centering
           \includegraphics[width=\textwidth]{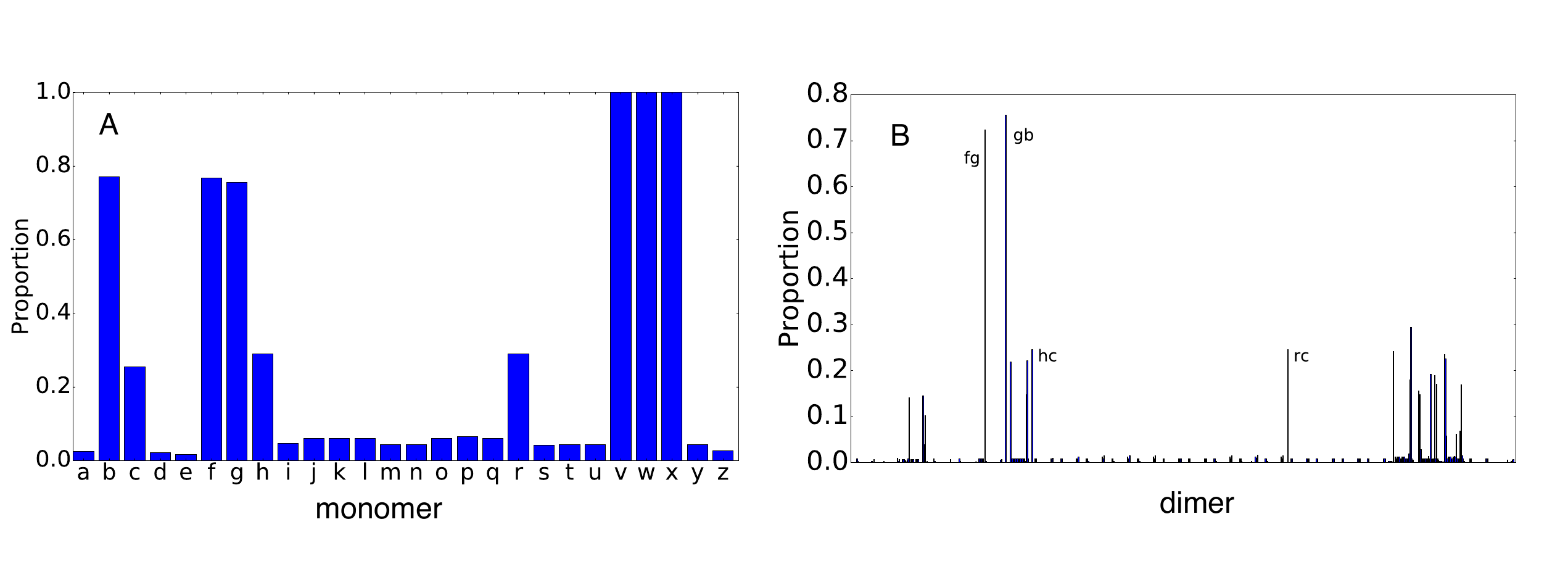}
    \caption{A: Distribution of monomers/single instructions (i.e., proportion of self-replicators containing a given monomer). B: Distribution of dimers (pairs of instructions). Dimers are ordered lexicographically on the $x$-axis (the proportion of {\tt fg}, {\tt gb}, {\tt rc}, and {\tt hc} dimers are labeled.) }
    \label{fig:dist}
\end{figure}

To study the genetic structure of these replicators, we obtained the distribution of instructions (monomers) across the replicators' genomes (Fig.~\ref{fig:dist}a). This distribution is biased, as every single replicator contained at least the three instructions required for replication: h-copy, h-alloc, and h-divide (denoted by ${\tt v}$, ${\tt w}$, and ${\tt x}$, respectively, see the mapping between instructions and the letter mnemonic in Table 1 in the Appendix). In addition, 75\% of replicators have a ${\tt b}$ (nop-B), an ${\tt f}$ (if-label), and a ${\tt g}$ (mov-head) instruction, while 25\% have a ${\tt c}$ (nop-C), an ${\tt h}$ (jmp-head), and an ${\tt r}$ (swap) instruction in their sequence. We also analyzed the distribution of sequential instruction pairs (dimers) and found that while most dimers do not occur in any self-replicators, the dimers ${\tt fg}$ and ${\tt gb}$ occur in approximately 70\% of the replicators (Fig.~\ref{fig:dist}b) and are highly over-represented . Other dimers such as ${\tt rc}$, ${\tt hc}$, and dimers containing ${\tt f}$,${\tt g}$,${\tt b}$,${\tt c}$,${\tt v}$,${\tt w}$, and ${\tt x}$ occur in approximately 20\%-30\% of replicators. 
%Fig. 1
 
\begin{figure}
        \includegraphics[width=\textwidth]{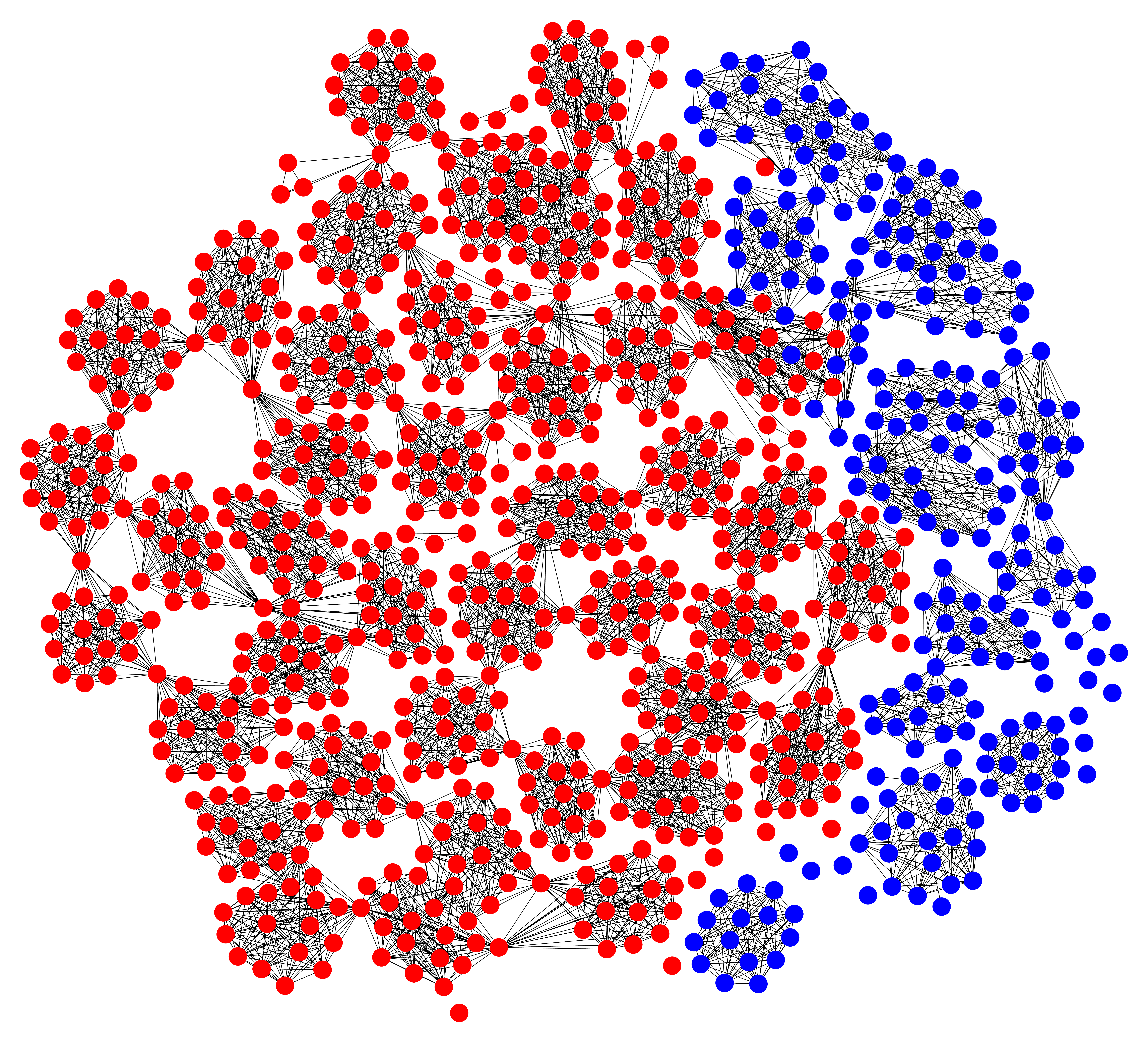}
	\caption{The complete fitness landscape of all 914 length-8 replicators. The replicators are colored by the class of motifs they contain ({\tt fg} replicators are colored in red, while {\tt hc} replicators are colored in blue.) The relative position between any pair of nodes reflects their distance in Hamming space, displayed via multi-dimensional scaling (MDS). As a consequence, it appears as if blue and red clusters are linked, which is not the case. One isolated fg-replicator (red) is close to an hc-replicator cluster (blue), but is not connected to it. All visible edges are between nodes that have a Hamming distance of 1 (i.e.\ they are a point mutation away from each other). }
    \label{fig:motifs}
\end{figure}

If there were no constraint on the genetic architecture, we would expect self-replicators to be distributed uniformly across the fitness landscape. However, we found instead that self-replicators are not distributed uniformly in the landscape, but are grouped into 41 distinct genotype clusters, shown in Fig.~\ref{fig:motifs}.

\begin{figure}[htbp] %  figure placement: here, top, bottom, or page
   \centering
   \includegraphics[width=4in]{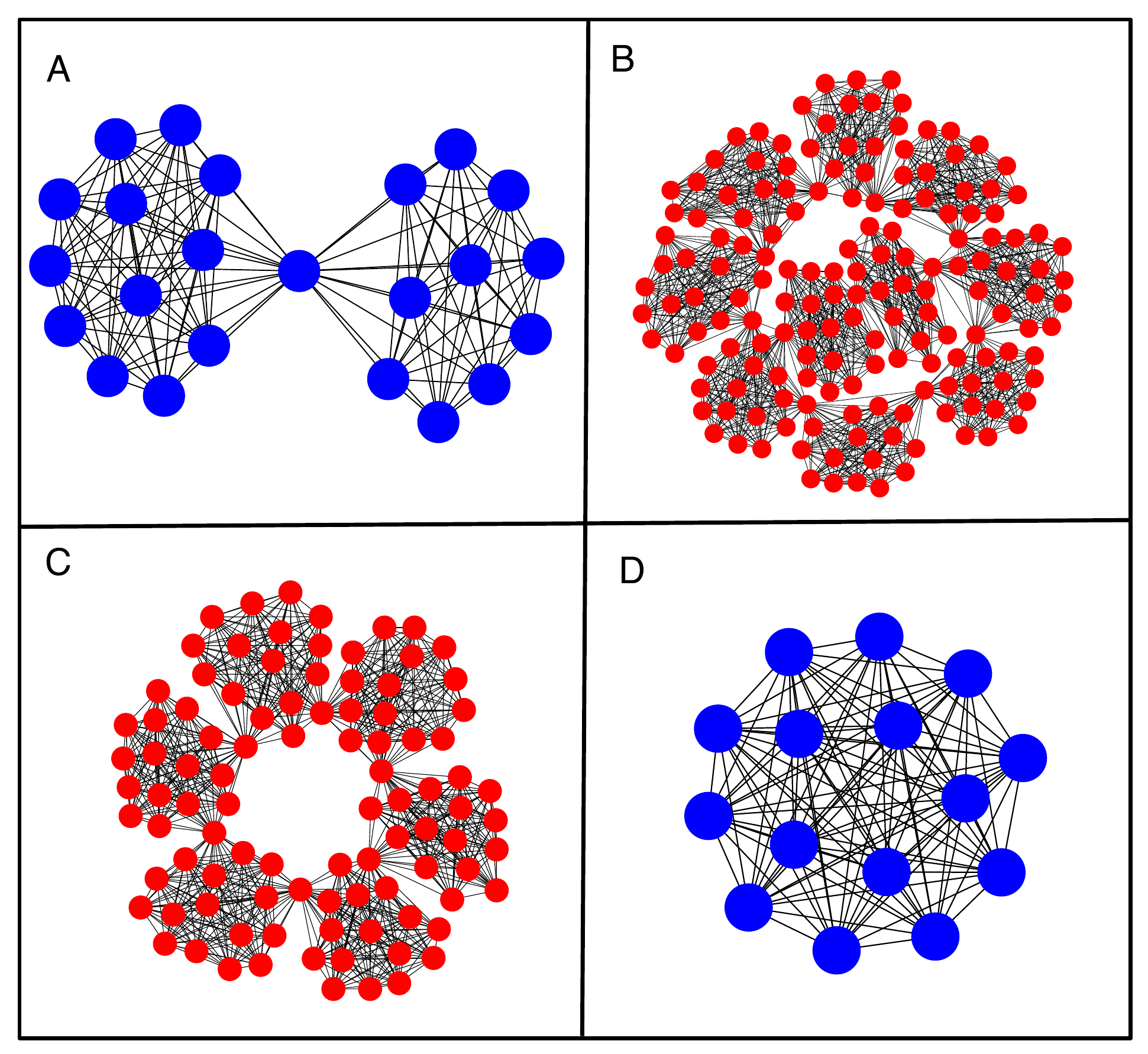} 
   \caption{Four clusters from the full landscape of self-replicators of $L=8$. A: A 23-node cluster of hc-replicators, B: the third-largest cluster in the network: an fg-replicator cluster with 165 members.  C: Another large fg-replicator cluster with 96 genotypes. D: A 15-node hc-replicator cluster. }
   \label{fig:clusters}
\end{figure}

The dimer distribution function we analyzed above separates primordial self-replicators into two major categories: those that carry {\tt fg/gb} motifs (``fg-replicators" for short), as opposed to those carrying {\tt hc/rc} motifs (``hc-replicators") instead. This separation into two classes was noted earlier from a smaller sample of the landscape \cite{labar2015does,labar2015evolvability}, which we corroborate here. By scanning the entire landscape we can confirm that these two types are the only types of self-replicators in the landscape, and the clusters of genotypes are homogeneous in the sense that fg-replicators and hc-replicators do not intermix (Fig.~\ref{fig:motifs}).
Fig.~\ref{fig:clusters} shows four examples of clusters pulled from the landscape, showing that they are tightly interconnected.

%Fig. 2

Many self-replicators are isolated and 20 of these clusters consist of only 1 genotype. However, most self-replicators are located in large clusters. Almost 75\% of the self-replicators are located in four major clusters with 212, 199, 165, and 95 genotypes each, and almost 96\% are contained within the 13 clusters that have at least 14 members. 
There is thus a distinct gap in the cluster size distribution, with small clusters ranging from 1-3 connected members, while the next largest size class is 14. 
%Fig. 3

\begin{figure}
        \includegraphics[width=0.8\textwidth]{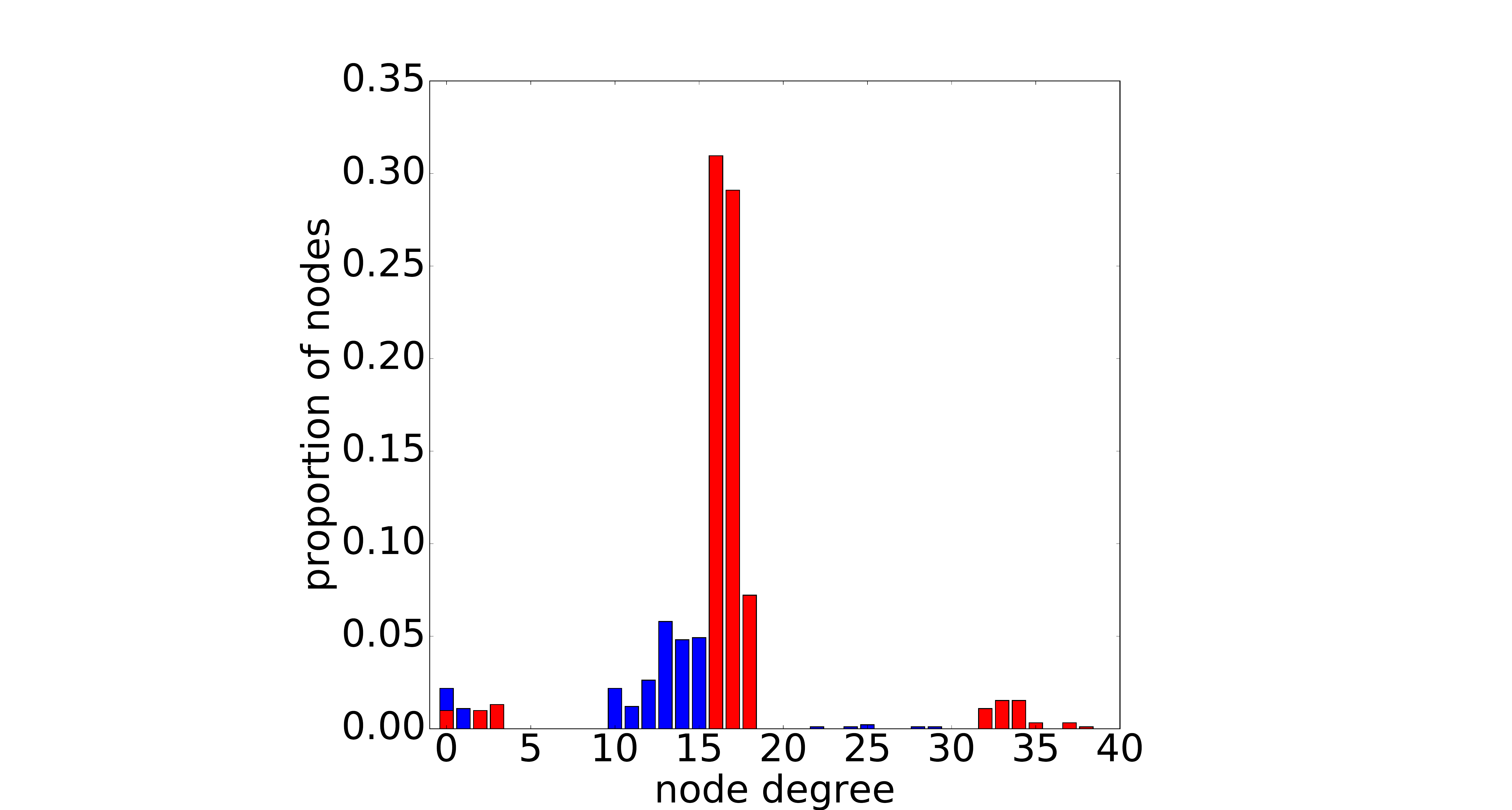}
	\caption{Edge distribution of all replicators in the fitness landscape of $L=8$. As each cluster has a particular edge distribution, the distributions of the two different kinds of replicators (fg-types and hc-types) do not overlap. Red: fg-replicators, blue: hc-replicators }
    \label{fig:edge_dist}
\end{figure}

We find that clusters of replicators are highly connected among each other, with a degree distribution that is sharply peaked around the mean degree of a cluster (see Fig.~\ref{fig:edge_dist}), which is similar to what is seen in neutral networks of random RNA structures~\cite{Aguirreetal2011}. We find that fg-replicators form the denser clusters. 
%Fig. 4

\begin{figure}
    \includegraphics[width=\textwidth]{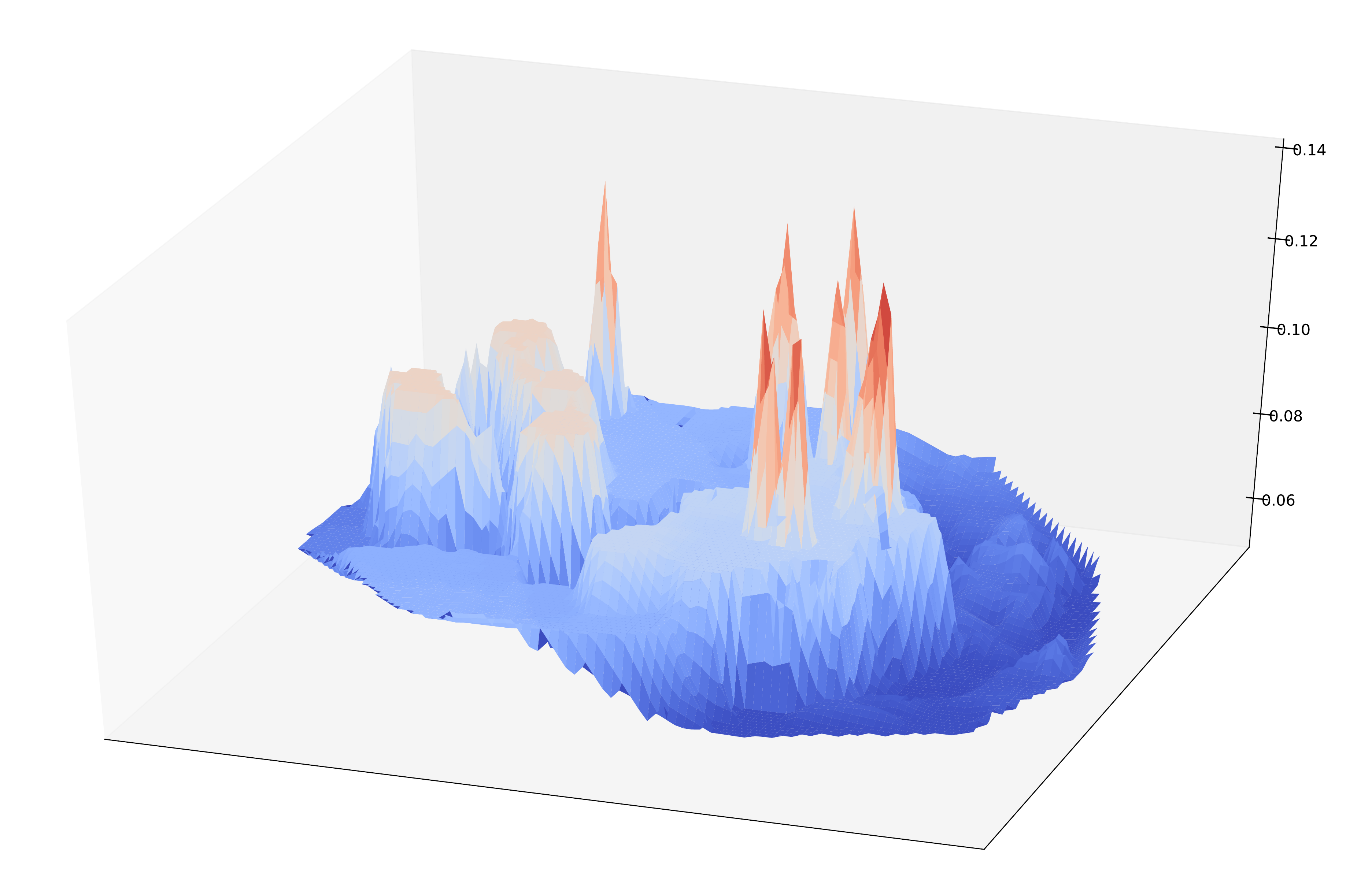}
    \caption{Ancestral fitness of all primordial self-replicators of $L=8$, where x-y coordinates are the same as the network in Fig \ref{fig:motifs}.}
    \label{fig:3dplot}
\end{figure}

The 914 self-replicators we found vary in fitness, but consistently we find that the fittest self-replicators contain the {\tt fg/gb} motifs and many of the lowest fitness self-replicators contain the {\tt hc/rc} motifs. In Fig.~\ref{fig:3dplot} we show the fitness as a function of the MDS-coordinate. In that figure, color denotes fitness according to the scale on the right. The highest peaks and plateaus all belong to fg-replicators. The hc-replicators appear as a valley (dark blue) bordering the group of fg-replicators.
%Fig.5 

\subsection*{Self-Replicator Evolvability }
\begin{figure}
    \centering
           \includegraphics[width=\textwidth]{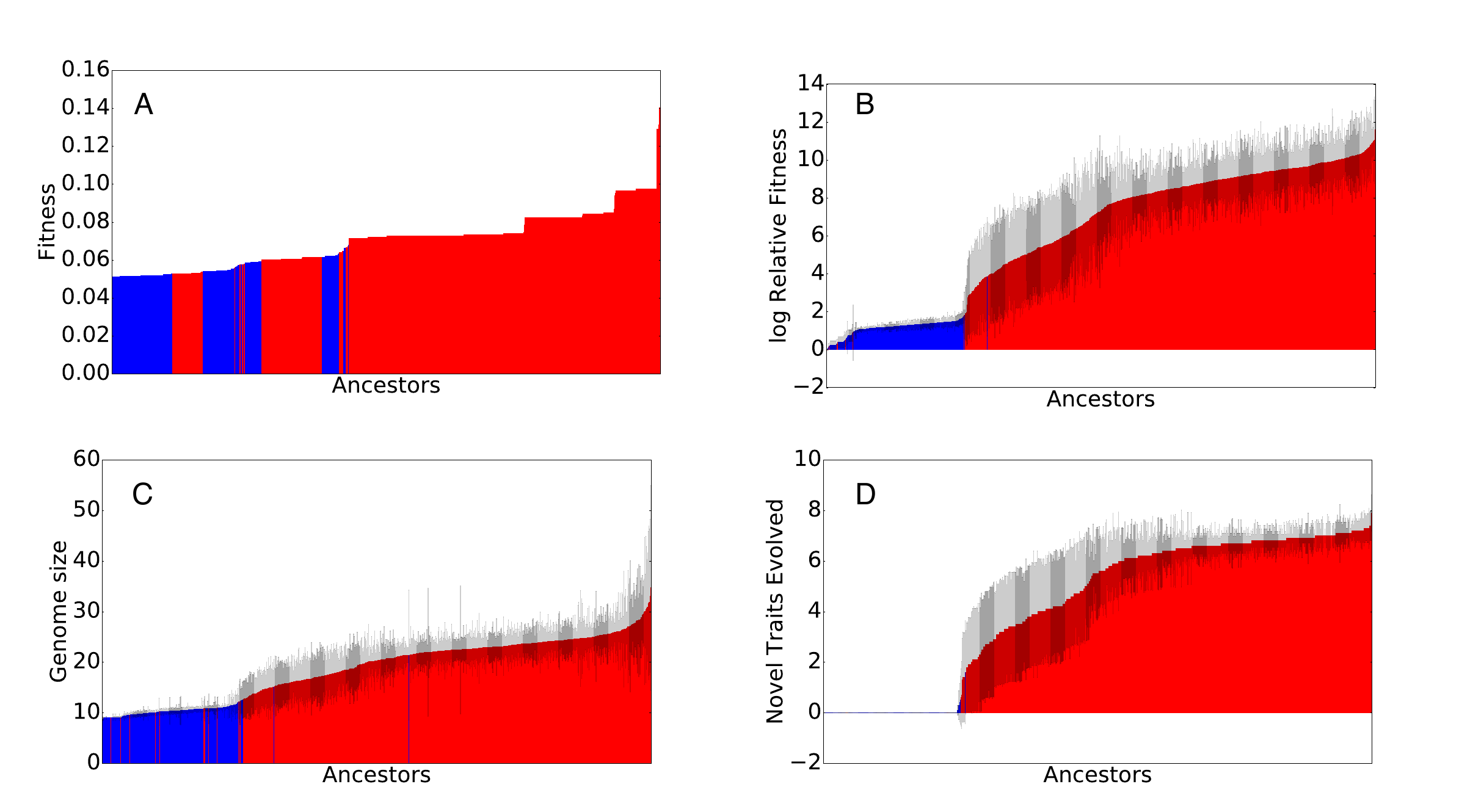}
   \caption{ Fitness and other characteristics of all $L=8$ self-replicators before and after evolution. A: Ancestral fitness of all replicators. B: Log mean relative fitness after $2 \times 10^4$ updates of evolution. C: Final genome size after $2 \times 10^4$ updates of evolution. D: Number of evolved traits after $2 \times 10^4$ updates of evolution.  In all plots, fg-replicators are in red and hc-replicators are in blue. Error bars (black) are twice the standard error of the mean. All plots are sorted in increasing order.}\label{fig:evol}
\end{figure}

In order to explore the subsequent role of historical contingency after the emergence of life, we tested the evolvability of all 914 self-replicators. First, we evolved each replicator separately. Almost all self-replicators could evolve increased fitness (Fig.~\ref{fig:evol}B). However, there was a wide range of mean relative fitness; fg-replicators clearly undergo more adaptation than hc-replicators. To explain why fg-replicators were more evolvable, we first looked at the evolution of genome size. Replicators with the {\tt fg/gb} motifs grew larger genomes than replicators with the {\tt hc/rc} motifs (Fig.~\ref{fig:evol}c). As larger genomes can allow for the evolution of novel traits in Avida, and thus fitness increases, we next checked whether the fg-replicators had evolved more {\em computational traits} than the hc-replicators. In Avida, traits are snippets of code that allow the avidian to gain energy from the environment, by performing logic operations on binary numbers that the environment provides (see Methods). Replicators with the {\tt fg/gb} motifs did evolve more novel traits than replicators with the {\tt hc/rc} motifs (Fig.~\ref{fig:evol}D). In fact, only fg-replicators evolved traits in these experiments.
%Fig. 6
Finally, we looked at the effect of historical contingency when all 914 replicators were competed against each other in one population. After 50,000 updates, we identify the most abundant genotype in 200 replicate experiments and reconstruct the line-of-descent  to determine which of the replicators gave rise to it (we call that replicator the ``progenitor").

\begin{figure}
    \includegraphics[width=\textwidth]{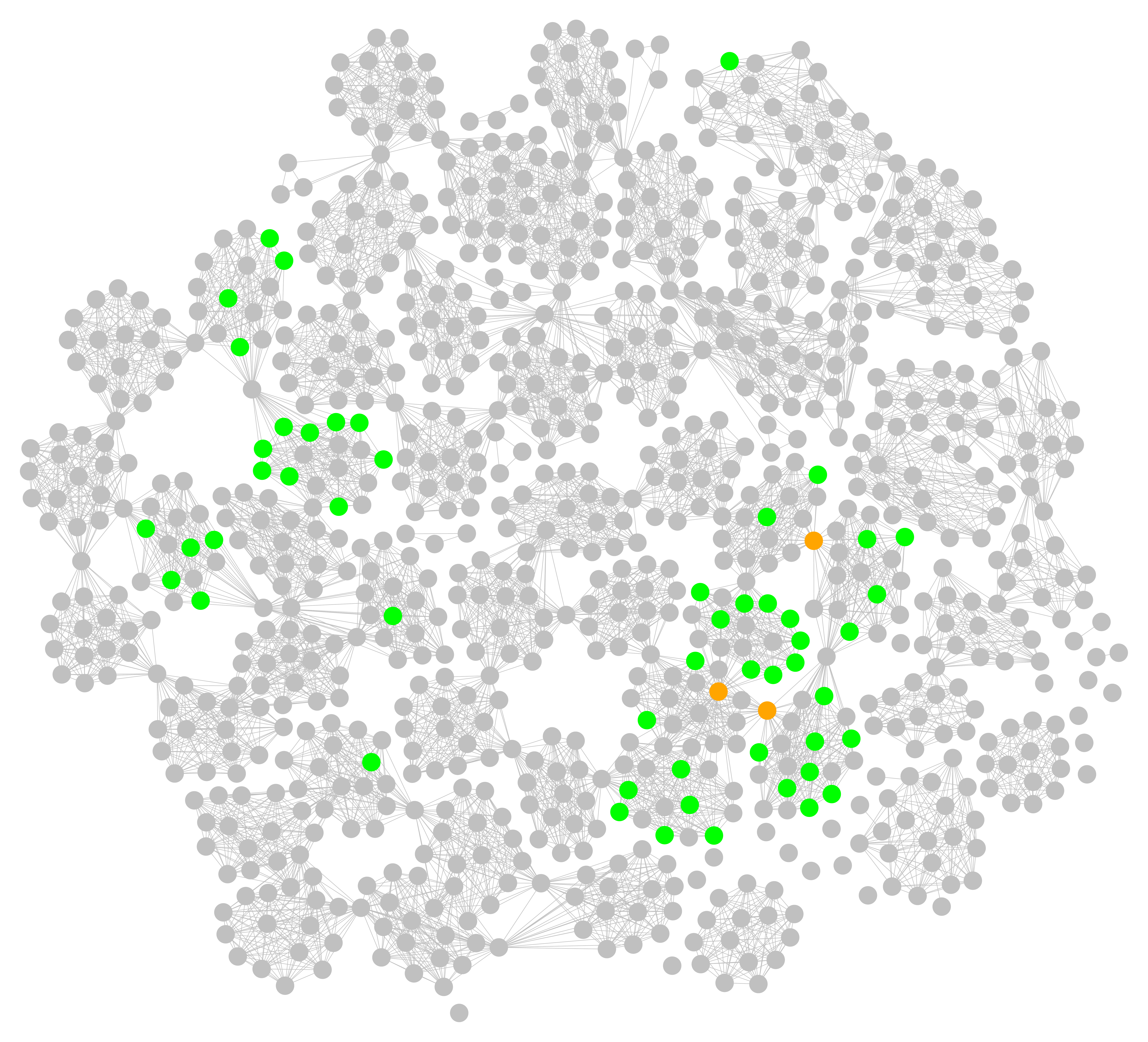}
    \caption{Location of ``progenitors" (ancestral types that were the origin of an evolved population 50,000 updates later) in the primordial landscape. Replicators that were never the ancestor genotype of the entire population are in grey. Those that outcompete all other genotypes in fewer than 6 (out of 200) competitions are colored in green. The three genomes that eventually become the ancestor of life in over 130 competitions are in orange. }
    \label{fig:soup}
\end{figure}

Most replicators did not emerge as the progenitor of life in these experiments (Fig.~\ref{fig:soup}). Three genotypes, {\tt vvwfgxgb}, {\tt vwvfgxgb}, and {\tt wvvfgxgb}, outcompete the other genotypes in 37, 49, and 45 populations out of 200, respectively, or in about 65\% of the competitions. The other progenitors of life were not distributed randomly among the other self-replicators either; most of them were present in the same clusters as the three genotypes from above.Thus, while history is a factor in which of the replicators becomes the seed of all life in these experiments, more than half the time the progenitor is one of the three highest-fitness sequences. Thus, life predominantly originates from the highest peaks of the primordial landscape.
%Fig. 7
    
\section*{Discussion}

Here, we tested the role of fitness landscape structure and historical contingency in the origin of self-replication in the digital evolution system Avida. We characterized the complete fitness landscape of all minimal-genome self-replicators and found that viable genotypes form clusters in the fitness landscape. These self-replicators can be separated into two replication classes, as we previously found for self-replicators with larger genomes~\cite{labar2015evolvability}. We also found that one of these replication classes (the fg-replicators) is more evolvable than the other, although the evolvability of each genotype varies. Finally, we show that, when all self-replicators are competed against each other in a digital ``primordial soup", three genotypes win over 65\% of the competitions and many of the other ``winners" come from the same genotype cluster. 

In a previous study with Avida, we found that 6 out of $10^9$ spontaneously-emergent genomes with 8 instructions could self-replicate~\cite{adami2015entropy}. Here, we found that 914 out of $\approx 2.8 \times 10^{11}$ genomes could replicate, consistent with our previous results. This concordance suggests that the information-theoretic theory of the emergence of life, originally proposed by Adami~\cite{adami2015information} and tested with Avida by Adami and LaBar~\cite{adami2015entropy}, can accurately explain the likelihood of the chance emergence of life. Thus, the emergence of self-replication, and life is dependent on the information required for such life.

By enumerating all of the length-8 self-replicators, we were able to show that self-replicators are not uniformly distributed across the fitness landscape and that viable genotypes cluster together. The size of these clusters varies: there are few clusters with many genotypes and many clusters with few genotypes, but the cluster size distribution  
has a gap. The edge distribution of the clusters is similar to what has been found in random RNA structures, and the mean degree differs between replicator types.

Genotypes with different replication mechanisms were in different clusters with no evolutionary trajectory between the two. Empirical studies of RNA-based fitness landscapes, biochemical model systems for the origin of life, also show that these landscapes consist of isolated fitness peaks with many non-viable genotypes~\cite{jimenez2013comprehensive,petrie2014limits}. The fact that both RNA-based landscapes~\cite{jimenez2013comprehensive,petrie2014limits} and these digital landscapes have similar structures suggests that the evolutionary patterns we see in these Avida experiments may be similar to those one would have seen in the origin of life on Earth. The presence of isolated genotype clusters in both digital and RNA fitness landscapes further suggests that the identity of the first self-replicator may determine life's future evolution, as other evolutionary trajectories are not accessible. However, if populations can evolve larger genomes, non-accessible evolutionary trajectories may later become accessible, as mathematical results on the structure of high-dimensional fitness landscapes suggest~\cite{gavrilets1997evolution}. 

To test for the effects of historical contingency in the origin of self-replication in Avida, we evolved all of the 914 replicators in an environment where they could increase in genome size and evolve novel traits. Previously, we found that the evolvability of spontaneously-emergent self-replicators varied and was determined by their replication mechanism~\cite{labar2015evolvability}. However, those genotypes possessed fixed-length genomes of 15 instructions. Here, we confirmed that the genotype of the first self-replicator, and more specifically the replication mechanism of the first replicator, determine the future evolution of novel traits in Avida. The fg-replicators showed high rates of trait evolution, while hc-replicators failed to evolve novel traits in most populations. However, we did not detect any trade-off in evolvability, as we previously found~\cite{labar2015evolvability}. This difference is likely due to their differences in capacity to increase in genome size, as genome size increases enhance the evolution of novel traits and fitness increases in Avida ~\cite{gupta2016evolution,labar2016different}. Would a similar dynamic occur in a hypothetical population of RNA-based replicators? While experimental evolution of RNA replicators has been performed, the selective environments resulted in genome size decreases~\cite{mills1967extracellular}. It is unknown how simple RNA replicators vary in their evolvability.

We also performed experiments to test for the role of historical contingency in scenarios where any self-replicator could become the progenitor of digital life. Here, we found that only three self-replicators (or their neighbors in the fitness landscape) became the last common ancestor in the majority of populations. This suggests a lack of contingency in the ancestral self-replicator, but emphasizes the role of the ancestral genotype in determining its future evolution. If life emerges rarely, then its future evolution will be determined by the specific genotype that first emerges, as shown from our first set of evolvability experiments (Fig.~\ref{fig:evol}). However, if simple self-replicators emerge frequently, then the future evolution is determined by the evolvability of the fittest replicators, a sort of clonal interference~\cite{gerrish1998fate} among possible progenitors of life. In this case, the self-replicators that most successfully invaded the population happened to also be of the type that evolved the largest genomes and most complex traits. However, it can be imagined that the opposite trend could occur~\cite{labar2015evolvability}, and then the progenitor of life would limit the future evolution of biological complexity.

\section*{Conclusions}
In this work we have performed the first complete mapping of a primordial sequence landscape in which replicators are extremely rare (about one replicator per 200 million sequences) and found two functionally inequivalent classes of replicators that differ in their fitness as well as evolvability, and that form distinct (mutationally disconnected) clusters in sequence space. In direct evolutionary competition, only the highest-fitness sequences manage to repeatedly become the common ancestor of all life in this microcosm, showing that despite significant diversity of replicators, historical contingency plays only a minor role during early evolution. 

While it is unclear how the results we obtained in this digital microcosm generalize to a biochemical microcosms, we are confident that they can guide our thinking about primordial fitness landscapes. The functional sequences we discovered here are extremely rare, but likely not as rare as putative biochemical primordial replicators. However, from a purely statistical point of view, it is unlikely that a primordial landscape consisting of sequences that are several orders of magnitude more rare would look qualitatively different, nor would we expect our results concerning historical contingency to change significantly. After all, random functional RNA sequences (but not replicators, of course) within a computational world~\cite{Aguirreetal2011}, chosen only for their ability to fold, show similar clustering and degree distributions as we find here. Follow-up experiments in the much larger $L=9$ landscape (currently under way) will reveal which aspects of the landscape are specific, and which ones are germane, in this digital microcosm.  

A comparison between fitness landscapes across a variety of evolutionary systems, both digital~\cite{pargellis2016digital} and biochemical~\cite{jimenez2013comprehensive}, will further elucidate commonalities expected for simple self-replicators. As the landscapes for these simple self-replicators are mapped, we expect general properties of primordial fitness landscapes to emerge, regardless of the nature of the replicator. As long as primordial self-replicators anywhere in the universe consist of linear heteropolymers that encode the information necessary to replicate, studies with digital microcosms can give us clues about the origin of life that experiments with terrestrian biochemistry cannot deliver.  

\section*{Acknowledgements}
This work was supported in part by the National Science Foundation's BEACON Center for the Study of Evolution in Action under Cooperative Agreement DBI-0939454. We wish to acknowledge the support of the Michigan State University High Performance Computing Center and the Institute for Cyber Enabled Research (iCER)
\appendix
\section*{Appendix}
\begin{table}[htbp]
   \centering
   \topcaption{Instruction set of the avidian programming language used in this study. The notation ?BX? implies that the command operates on a register specified by the subsequent nop instruction (for example, nop-A specifies the AX register, and so forth). If no nop instruction follows, use the register BX as a default. More details about this instruction set can be found in~\cite{ofria2009avida}.} % requires the topcapt package
   \begin{tabular}{@{} lll @{}} % Column formatting, @{} suppresses leading/trailing space
      \toprule
      Instruction    & Description & Symbol\\
      \midrule
nop-A    & no operation (type A) & a \\
nop-B   & no operation (type B) & b \\
nop-C   & no operation (type C) & c \\
if-n-equ & Execute next instruction only-if ?BX? does not equal complement & d\\
if-less & Execute next instruction only if ?BX? is less than its complement & e\\
if-label & Execute next instruction only if template complement was just copied & f\\
mov-head & Move instruction pointer to same position as flow-head & g\\
jmp-head & Move instruction pointer by fixed amount found in register CX & h\\
get-head & Write position of instruction pointer into register CX & i\\
set-flow & Move the flow-head to the memory position specified by ?CX? & j\\
shift-r & Shift all the bits in ?BX? one to the right & k\\
shift-l & Shift all the bits in ?BX? one to the left & l\\
inc & Increment ?BX? & m\\
dec & Decrement ?BX? & n\\
push &       Copy value of ?BX? onto top of  current stack & o\\
pop & Remove number from current stack and place in ?BX? & p\\
swap-stk & Toggle the active stack & q\\
swap & Swap the contents of ?BX? with its complement & r\\
add & Calculate  sum of BX and CX; put  result in ?BX? & s\\
sub & Calculate  BX minus CX; put result in ?BX? & t\\
nand & Perform bitwise NAND on BX and CX; put  result in ?BX? & u\\
h-copy &  Copy instruction from read-head to write-head and advance both & v\\
h-alloc & Allocate memory for offspring & w\\
h-divide & Divide off an offspring located between read-head and write-head & x \\
IO &  Output value ?BX? and replace with new input & y\\
h-search & Find complement template and place flow-head after it & z\\
      \bottomrule
   \end{tabular}
  % \caption{}
   \label{avidainst}
\end{table}

%\bibliographystyle{rsta}
%\bibliography{citations}

\end{document}